\documentclass[preprint,aps,showpacs]{revtex4}
\usepackage{amsmath}
\usepackage{amssymb}
\usepackage{bm}
\usepackage{graphics}
\usepackage[dvips]{graphicx,color}
\usepackage{dcolumn}
\def\lesssim{\ \raise.3ex\hbox{$<$}\kern-0.8em\lower.7ex\hbox{$\sim$}\ }
\def\gesim{\ \raise.3ex\hbox{$>$}\kern-0.8em\lower.7ex\hbox{$\sim$}\ }

\begin{document}
\title{Triplet pair amplitude in a trapped $s$-wave superfluid Fermi gas with broken spin rotation symmetry. II. Three dimensional continuum case}
\author{Daisuke Inotani, Ryo Hanai, and Yoji Ohashi}
\affiliation{Department of Physics, Faculty of Science and Technology, Keio University, 3-14-1, Hiyoshi, Kohoku-ku, Yokohama 223-8522, Japan}
\date{\today}
\begin{abstract}
We extend our recent work [Y. Endo et. al., Phys. Rev. {\bf 92}, 023610 (2015)] for a parity-mixing effect in a model two-dimensional lattice fermions to a {\it realistic three-dimensional} ultracold Fermi gas. Including effects of broken local spatial inversion symmetry by a trap potential within the framework of the real-space Bogoliubov-de Gennes theory at $T=0$, we point out that an {\it odd-parity} $p$-wave Cooper-pair amplitude is expected to have already been realized in previous experiments on an (even-parity) $s$-wave superfluid Fermi gas with spin imbalance. This indicates that, when one suddenly changes the $s$-wave pairing interaction to an appropriate $p$-wave one by using a Feshbach technique in this case, a non-vanishing $p$-wave superfluid order parameter is immediately obtained, which is given by the product of the $p$-wave interaction and the $p$-wave pair amplitude that has already been induced in the spin-imbalanced $s$-wave superfluid Fermi gas. Thus, by definition, the system is in the $p$-wave superfluid state, at least just after this manipulation. Since the achievement of a $p$-wave superfluid state is one of the most exciting challenges in cold Fermi gas physics, our results may provide an alternative approach to this unconventional pairing state. In addition, since the parity-mixing effect cannot be explained as far as one deals with a trap potential in the local density approximation (LDA), it is considered as a crucial example which requires us to go beyond LDA.
\end{abstract}
\maketitle
\par
\section{Introduction}
\par
In a recent paper \cite{Endo}, using symmetry considerations, we pointed out that the spatial inhomogeneity caused by a harmonic trap potential, which breaks the local spatial inversion symmetry of the system except at the trap center, may induce an odd-parity (spin-triplet) Cooper-pair amplitude  in the even-parity (spin-singlet) $s$-wave Fermi superfluid, when the spin rotation symmetry is also broken. Here, the local spatial inversion symmetry is the inversion symmetry of the system with respect to an inversion center. For example, when we carry out this symmetry operation for a one-dimensional harmonic potential $V(x)=cx^2$ ($c>0$) with respect to $x=a$, one finds the breakdown of this symmetry as, $V(x) \to V(-x+2a) \neq V(x)$, unless $a=0$ (trap center). On the other hand, the spin rotation symmetry in the present case is the symmetry with respect to exchange of two pseudospins $\uparrow \leftrightarrows \downarrow$. In Ref. \cite{Endo}, to confirm the above-mentioned parity-mixing effect in a simple manner, we explicitly evaluated the induced odd-parity pair amplitude in a toy two-dimensional lattice fermions satisfying the above two conditions.
\par
The above-mentioned work \cite{Endo} is strongly motivated by the current status of cold Fermi gas physics. That is, the $p$-wave superfluid phase transition has not been realized yet, in spite of great experimental efforts \cite{Regal2003,Ticknor2004,Salomon2004,Gunter2005,Ketterle2005,Gaebler2007,Mukaiyama2013}. One serious reason for this is that, while a $p$-wave pairing interaction is necessary to realize a $p$-wave superfluid Fermi gas, it also causes the three-body loss \cite{Guririe2007,Gurarie2008,Castin}, as well as dipolar relaxation \cite{Gaebler2007}, leading to very short lifetime of $p$-wave Cooper pairs. Because of this, when one tunes an external magnetic field to be close to a $p$-wave Feshbach resonance, although, theoretically, a strong $p$-wave pairing interaction is expected to give a high $p$-wave superfluid phase transition temperature \cite{Ho2005,Ohashi2005,Inotani2012,Inotani2015}, the $p$-wave pairs are actually destroyed before the $p$-wave condensate grows enough.
\par
However, if one can prepare a non-vanishing $p$-wave pair amplitude (which is symbolically written as $\langle\Psi_\sigma({\bm r})\Psi_{\sigma'}({\bm r}')\rangle$, where $\Psi_\sigma({\bm r})$ is the field operator of a fermion with spin-$\sigma$) {\it without} using a $p$-wave pairing interaction, this current situation would be improved to some extent, because one can then avoid the serious short lifetime problem of $p$-wave Cooper pairs. Of course, the system is still {\it not} in the $p$-wave superfluid state, because of the vanishing $p$-wave superfluid order parameter, 
\begin{equation}
\Delta_p({\bm r},{\bm r}')=U_p({\bm r},{\bm r}')\langle\Psi_\sigma({\bm r})\Psi_{\sigma'}({\bm r}')\rangle,
\label{eq1}
\end{equation}
due to the absence of a $p$-wave pairing interaction $U_p({\bm r},{\bm r}')$. In this case, however, when one rapidly introduces a $p$-wave pairing interaction $U_p({\bm r},{\bm r}')$ associated with a $p$-wave Feshbach resonance by adjusting an external magnetic field, a non-vanishing $p$-wave superfluid order parameter (which is given by the product of this interaction and the $p$-wave pair amplitude which has already been prepared before this manipulation) is immediately obtained. Then, by definition, the $p$-wave superfluid state which is characterized by the $p$-wave superfluid order parameter $\Delta_p({\bm r},{\bm r}')$ would be realized, at least just after the introduction of the $p$-wave interaction. Once the $p$-wave superfluid state is realized, as usual, the $p$-wave interaction would cause the particle loss, as well as dipolar relaxation. However, this idea might be alternative to the conventional approach where a $p$-wave interaction is used {\it from the beginning}.
\par
The purpose of this paper is to assess to what extent the parity-mixing effect discussed in our recent paper \cite{Endo} can be used to prepare a $p$-wave pair amplitude in the above idea. For this purpose, beyond the previous toy two-dimensional lattice model, we deal with a realistic three-dimensional {\it continuum} $s$-wave superfluid Fermi gas in a harmonic trap. This extension is very relevant, because all the current experiments on $s$-wave superfluid Fermi gases \cite{Regal,Zwierlein,Kinast,Bartenstein,Horikoshi} are done in the absence of deep optical lattice.
\par
Between the above-mentioned two conditions for the parity-mixing effect, the local spatial inversion symmetry is always broken in a harmonic trap, when one takes the inversion center away from the trap center. Thus, no additional setup is necessary for this first condition. However, we should note that the local density approximation (LDA), which has widely been used in considering a trapped Fermi gas \cite{Tsuchiya,Watanabe,Haussmann,Bulgac}, cannot correctly treat this symmetry breaking. This is simply because LDA deals with the system at each spatial position as a uniform system with a position dependent Fermi chemical potential involving effects of a trap potential. Thus, in this paper, we employ the real-space Bogoliubov-de Gennes mean-field theory \cite{deGennes,Ohashitrap}, to fully take into account the spatial inhomogeneity in a trap. 
\par
For the second required condition for the parity-mixing effect (broken spin rotation symmetry), a spin-imbalanced $s$-wave Fermi superfluid, which has already been realized in $^6$Li Fermi gases \cite{Zwierlein2006,Partridge,Shin,Kashimura}, is most promising, where the density difference between two Fermi species naturally breaks this symmetry. In a trapped $s$-wave superfluid Fermi gas with spin imbalance, we evaluate how large the $p$-wave Cooper-pair amplitude is induced, to assess whether or not this phenomenon can be practically used for our purpose toward the realization of a $p$-wave superfluid Fermi gas. In this paper, we also investigate the other two cases when the spin rotation symmetry is broken by (1) mass imbalance \cite{Taglieber2008,Wille2008,Spiegelhalder2009,Lin2006,Hanai2013,Hanai2014}, or (2) a spin-dependent trap potential.
\par
This paper is organized as follows. In Sec.II, we explain our real-space formulation. In Sec.III, we numerically evaluate the magnitude of the triplet pair amplitude induced in a trapped $s$-wave superfluid Fermi gas with spin imbalance. The cases with mass imbalance and a spin-dependent trap potential are also examined there. Throughout this paper, we take $\hbar=k_{\rm B}=1$, for simplicity.
\section{Formalism}
We consider a three-dimensional continuum $s$-wave superfluid Fermi gas in a harmonic trap potential, described by the Hamiltonian
\begin{eqnarray}
H&=&\sum_\sigma \int d {\bm r} \psi_{\sigma}^{\dagger}\left( {\bm r} \right)
\left[
-\frac{\nabla^2}{2m_\sigma}-\mu_\sigma+V_{\sigma}\left( r \right)
\right]
\psi_{\sigma}\left( {\bm r} \right)
\nonumber
\\
&-&
U_s \int d {\bm r} 
\psi_{\uparrow}^{\dagger}\left( {\bm r} \right)
\psi_{\downarrow}^{\dagger}\left( {\bm r} \right)
\psi_{\downarrow}\left( {\bm r} \right)
\psi_{\uparrow}\left( {\bm r} \right)
.
\label{eq2}
\end{eqnarray}
Here, $\psi_{\sigma} \left( \bm r \right)$ is the field operator of a Fermi atom at ${\bm r}$, with the pseudospin $\sigma = \uparrow, \downarrow$, the Fermi chemical potential $\mu_\sigma$, and the atomic mass $m_\sigma$. $-U_s$($<0$) is an $s$-wave pairing interaction. Fermi atoms with spin-$\sigma$ feel the harmonic potential $V_{\sigma}\left( {\bm r} \right) = m_\sigma \omega_\sigma^2 r^2/2$. For simplicity, we only consider an isotropic trap in this paper. To induce the triplet Cooper pairs, it is necessary to simultaneously break the spatial inversion symmetry and the spin rotation symmetry \cite{Endo}. The former is naturally realized by the harmonic trap potential $V_\sigma({\bm r})$, except at the trap center (${\bm r}=0$). For the latter, we mainly discuss the case with spin imbalance, which is realized by taking $\mu_\uparrow \neq \mu_\downarrow$ in Eq. (\ref{eq2}). In addition to this, we also investigate the cases with mass imbalance ($m_\uparrow \neq m_\downarrow$), as well as spin dependent trap potential ($\omega_\uparrow \neq \omega_\downarrow$), that also break the spin rotation symmetry.
\par
To fully take into account the broken local spatial inversion symmetry by a harmonic trap (which is essential for the parity-mixing phenomenon \cite{Endo}), we go beyond LDA, to employ the real-space Bogoliubov-de Gennes (BdG) mean-field theory \cite{deGennes,Ohashitrap}. In this case, the Hamiltonian Eq. (\ref{eq2}) is reduced to 
\begin{eqnarray}
H_{\rm {BdG}}&=&\sum_\sigma \int d {\bm r} \psi_{\sigma}^{\dagger}\left( {\bm r} \right)
\left[
-\frac{\nabla^2}{2m_\sigma}-\mu_\sigma+V_{\sigma}\left( r \right)
\right]
\psi_{\sigma}\left( {\bm r} \right)
\nonumber
\\
&-&
U_s\sum_{\sigma} \int d {\bm r}
n_{-\sigma} \left(r \right) 
\psi_{\sigma}^{\dagger}\left( {\bm r} \right)
\psi_{\sigma}\left( {\bm r} \right)
+\int d {\bm r} \left[
\Delta \left( r \right)
\psi_{\uparrow}^{\dagger}\left( {\bm r} \right)
\psi_{\downarrow}^{\dagger}\left( {\bm r} \right)
+h.c.
\right]
.
\label{eq3}
\end{eqnarray}
The BdG Hamiltonian in Eq. (\ref{eq3}) involves two mean-field parameters that are determined self-consistently, that is, one is the $s$-wave superfluid order parameter $\Delta({\bm r}) = -U_s \left\langle\psi_{\downarrow} \left( {\bm r} \right)\psi_{\uparrow} \left( {\bm r} \right) \right\rangle$, and the other is the Hartree potential $-U_sn_{-\sigma}(r)=-U_s\left\langle\psi^{\dagger}_{-\sigma} \left( {\bm r} \right)\psi_{-\sigma} \left( {\bm r} \right) \right\rangle$. For these self-consistent calculations, it is convenient to expand the fermion field operator  $\psi_\sigma\left( {\bm r} \right)$ with respect to the eigenfunctions $f^{\sigma}_{nlm} \left( {\bm r} \right)$ of the one-particle Schr\"odinger equation,
\begin{eqnarray}
\left[
-\frac{\nabla^2}{2m_\sigma}-\mu_\sigma+V_{\sigma}\left( r \right)
\right]f^{\sigma}_{nlm} \left( {\bm r} \right)=\xi_{nl}^\sigma f^{\sigma}_{nlm} \left( {\bm r} \right)
,
\label{eqbdg1}
\end{eqnarray}
as
\begin{eqnarray}
\psi_\sigma\left( {\bm r} \right)
=\sum_{n=0}^{\infty}\sum_{l=0}^{\infty}\sum_{m=-l}^{l} c_{nlm \sigma} f^{\sigma}_{nlm} \left( {\bm r} \right)
,
\label{eqbdg2}
\end{eqnarray}
where $c_{nlm \sigma}$ is the annihilation operator of a Fermi atom with the eigenenergy
\begin{eqnarray}
\xi_{nl}^{\sigma}=\omega_\sigma \left( 2 n + l +\frac{3}{2} \right)-\mu_\sigma
.
\label{eqbdg3}
\end{eqnarray}
The eigenfunction $f^{\sigma}_{nlm} \left( {\bm r} \right)$ can be written as $f^{\sigma}_{nlm}\left( {\bm r} \right)=R_{nl}^{\sigma} \left( r \right) Y_{lm}\left( \hat{{\bm r}} \right)$, where $Y_{lm}\left(\hat{\bm r} \right)$ is the spherical harmonics (where $\hat{\bm r}={\bm r}/r$), and the radial component $R_{nl}^{\sigma} \left( r \right)$ has the form,
\begin{eqnarray}
R_{nl}^{\sigma} \left( r \right)
=\sqrt{2}\left( m_\sigma \omega_\sigma \right)^\frac{3}{4}
\sqrt{\frac{n!}{\left( n+l +\frac{1}{2} \right)!}}
e^{-\frac{x_{\sigma}^2}{2}}
x_{\sigma}^{l}
L_n^{l+\frac{1}{2}}\left(x_{\sigma}^{2} \right)
,
\label{eq4}
\end{eqnarray}
with 
$x_{\sigma}=\sqrt{m_\sigma \omega_\sigma} r$ 
and 
$L_n^{l+\frac{1}{2}}\left(x \right)$ 
being the Laguerre polynomial. Substituting Eq. (\ref{eqbdg2}) into the BdG Hamiltonian in Eq. (\ref{eq3}), one has 
\begin{eqnarray}
H_{\rm BdG}=\sum_{l=0}^{\infty}\sum_{m=-l}^{l}  
C^{\dagger}_{lm}
\left(
\begin{array}{cc}
\xi^{\uparrow}_{nl}\delta_{nn'}
+J^{l,\uparrow}_{n n'} 
& \left( -1 \right)^{m} F^l_{n n'}  \\
\left( -1 \right)^{m} \left(F^l \right)^T_{n n'}
& -\xi^{\downarrow}_{nl}\delta_{nn'}
- J^{l, \downarrow}_{n n'} 
\end{array}
\right)
C_{lm}
,
\label{eq5}
\end{eqnarray}
where $
C_{lm}^{\dagger}=\{C_{lm,i}\}^{\dagger}
=\left(
c^{\dagger}_{0,l,m,\uparrow}, 
\cdots,
c^{\dagger}_{N_l,l,m,\uparrow},
c_{0,l,-m,\downarrow},
\cdots ,
c_{N_l,l,-m,\downarrow}
\right)
$. In Eq. (\ref{eq5}), we have implicitly introduced a high energy cutoff $\omega_{\rm c}^{\sigma}=\omega_\sigma \left( N_{\rm c}+3/2 \right)$ to eliminate the well-known ultraviolet divergence involved in the present BCS model, so that $N_l$ in $C_{lm}^{\dagger}$ is chosen as the maximum integer which satisfies $2N_l+l \le N_{\rm c}$ for a given value of $l$ ($\ge0$). The $\left( N_l+1 \right) \times \left( N_l+1 \right)$ matrices $\hat{F}^l$ and $\hat{J}^{l, \sigma}$ in Eq. (\ref{eq5}) are, respectively, the $s$-wave superfluid order parameter and the Hartree potential in the present basis set $\{ f^{\sigma}_{nlm} \left( {\bm r} \right) \}$, given by
\begin{eqnarray}
F^l_{nn'}&=&\int dr 
r^2 R^\uparrow_{nl} \left( r \right)
\Delta \left( r \right)
R^\downarrow_{n'l} \left( r \right)
,
\label{eq6}
\\
J^{l,\sigma}_{nn'}&=&-U_s\int dr 
r^2 R^\sigma_{nl} \left( r \right)
n_{-\sigma} \left( r \right)
R^\sigma_{n'l} \left( r \right)
.
\label{eq7}
\end{eqnarray}
As usual, we diagonalize Eq. (\ref{eq5}) by the Bogoliubov transformation 
\begin{eqnarray}
\gamma_{lmi}=\sum_{j=1}^{2\left( N_l+1 \right)} W_{ij}^{l}C_{lm,i}
,
\label{eqbdg4}
\end{eqnarray}
for each ($l,m$), which gives
\begin{eqnarray}
H_{\rm BdG}=\sum_{l=0}^{\infty}\sum_{m=-l}^{l} \sum_{i=1}^{2(N_l+1)} E^l_{i} \gamma_{lmi}^{\dagger}\gamma_{lmi}
.
\label{eq8}
\end{eqnarray}
Here, $E^{l}_{i}$ describes the Bogoliubov single-particle excitations. We briefly note that this eigenenergy is independent of $m$ because of the rotation symmetry of the system with respect to the trap center. 
\par
In this paper, we numerically carry out the Bogoliubov transformation, to determine $W_{ij}^{l}$ and $E_i^{l}$. Then, the self-consistent equations for $\Delta(r)$ and $n_\sigma(r)$ are obtained as
\begin{eqnarray}
\Delta(r)&=&U_s \sum_{lnn'} \frac{2l +1}{4\pi} 
R^{\downarrow}_{nl}(r) R^{\uparrow}_{n'l}(r)d^{l}_{nn'}
,
\label{eq9}
\\ 
n_\sigma(r)&=&\sum_{lnn'} \frac{2l +1}{4\pi} 
R^{\sigma}_{nl}(r) R^{\sigma}_{n'l}(r)
\eta^\sigma_{lnn'}
,
\label{eq10}
\end{eqnarray}
where
\begin{eqnarray}
d_{lnn'}&=&\sum_{i=1}^{2(N_l+1)} W^{l}_{n+N_l+1,i}W^{l}_{n',i} f \left( E^l_{i} \right)
,
\label{eq12}
\\ 
\eta^\uparrow_{lnn'}&=&\sum_{i=1}^{2(N_l+1)} W^{l}_{n,i}W^{l}_{n',i} f \left( E^l_{i} \right)
,
\label{eq13}
\\ 
\eta^\downarrow_{lnn'}&=&\sum_{i=1}^{2(N_l+1)} W^{l}_{n+N_l+1,i}W^{l}_{n'+N_l+1,i} \left[ 1- f \left( E^l_{i} \right) \right]
.
\label{eq14}
\end{eqnarray}
In Eqs. (\ref{eq12})-(\ref{eq14}), $f(\varepsilon)$ is the Fermi distribution function. We actually (numerically) solve Eqs. (\ref{eq9}) and (\ref{eq10}), together with the number equation,
\begin{eqnarray}
N_\sigma=\int d{\bm r} n_{\sigma} ({\bm r})
,
\label{numeq}
\end{eqnarray}
to determine $\Delta(r)$, $n_\sigma(r)$ and $\mu_\sigma$ in a consistent manner.
\par
The spin-singlet Cooper-pair amplitude ($\Phi_{\rm S}$), as well as the spin-triplet ones ($\Phi_{\rm T}^{S_z=0,\pm1}$), is given by, respectively,
\begin{eqnarray}
\Phi_{\rm{S}} \left( {\bm R},  {\bm r}_{\rm rel} \right)
&=&\frac{1}{\sqrt{2}}
\left[
\left\langle
\psi_\uparrow\left( {\bm R}+\frac{{\bm r}_{\rm rel}}{2} \right)
\psi_\downarrow\left( {\bm R}-\frac{{\bm r}_{\rm rel}}{2} \right)
\right\rangle
-
\left\langle
\psi_\downarrow\left( {\bm R}+\frac{{\bm r}_{\rm rel}}{2} \right)
\psi_\uparrow\left( {\bm R}-\frac{{\bm r}_{\rm rel}}{2} \right)
\right\rangle
\right]
,
\nonumber
\\
\label{eq15}
\\
\Phi_{\rm{T}}^{S_z=1} \left( {\bm R},  {\bm r}_{\rm rel} \right)
&=&
\left\langle
\psi_\uparrow\left( {\bm R}+\frac{{\bm r}_{\rm rel}}{2} \right)
\psi_\uparrow\left( {\bm R}-\frac{{\bm r}_{\rm rel}}{2} \right)
\right\rangle
,
\label{eq16}
\\
\Phi_{\rm{T}}^{S_z=0} \left( {\bm R},  {\bm r}_{\rm rel} \right)
&=&\frac{1}{\sqrt{2}}
\left[
\left\langle
\psi_\uparrow\left( {\bm R}+\frac{{\bm r}_{\rm rel}}{2} \right)
\psi_\downarrow\left( {\bm R}-\frac{{\bm r}_{\rm rel}}{2} \right)
\right\rangle
+
\left\langle
\psi_\downarrow\left( {\bm R}+\frac{{\bm r}_{\rm rel}}{2} \right)
\psi_\uparrow\left( {\bm R}-\frac{{\bm r}_{\rm rel}}{2} \right)
\right\rangle
\right]
,
\nonumber
\\
\label{eq17}
\\
\Phi_{\rm{T}}^{S_z=-1} \left( {\bm R},  {\bm r}_{\rm rel} \right)
&=&
\left\langle
\psi_\downarrow\left( {\bm R}+\frac{{\bm r}_{\rm rel}}{2} \right)
\psi_\downarrow\left( {\bm R}-\frac{{\bm r}_{\rm rel}}{2} \right)
\right\rangle
,
\label{eq18}
\end{eqnarray}
where ${\bm R}$ and ${\bm r}_{\rm rel}$ are the center of mass position of pairs and the relative coordinate between two atoms forming a Cooper pair, respectively. The superscript $S_z=0, \pm 1$ means the $z$ component of the total spin of a triplet Cooper pair. In the present $s$-wave superfluid phase, noting that
\begin{eqnarray}
\Delta(r)=-U_s \left\langle\psi_{\downarrow} \left( {\bm r} \right)\psi_{\uparrow} \left( {\bm r} \right) \right\rangle
=\sqrt{2}U_s \Phi_{\rm S} \left( {\bm r}, 0 \right)
,
\label{eqdelta}
\end{eqnarray}
we find that the spin-singlet pair amplitude $\Phi_{\rm S} \left( {\bm R}, {\bm r}_{\rm {rel}} \right)$ is always non-vanishing, at least $|{\bm r}_{\rm {rel}}|=0$. On the other hand, among the three triplet components in Eqs. (\ref{eq16})-(\ref{eq18}), we find from the symmetry consideration \cite{Endo} that only $\Phi_{\rm T}^{S_z=0}$ may be non-vanishing in the present case. In this regard, we briefly note that this is difference from the case with a synthetic spin-orbit coupling \cite{revSOC,Rb-SOC1,Rb-SOC2,Rb-SOC3,Rb-SOC4,Li-SOC,K-SOC,coldSOC5,coldSOC6}, where all the three components $\Phi_{\rm T}^{S_z=0,\pm 1}$ may be non-vanishing \cite{Yamaguchi,Hu2011,2B1}. In the following, we simply write the $S_z=0$ component as $\Phi_{\rm T}\left( {\bm R}, {\bm r}_{\rm {rel}} \right)$.
\par
Substituting Eq. (\ref{eqbdg2}) into Eqs. (\ref{eq15}) and (\ref{eq17}), we have
\begin{eqnarray}
\Phi_{\rm S} \left( {\bm R}, {\rm r}_{\rm rel} \right)
&=&-\frac{1}{\sqrt{2}}\sum_{lnn'} 
\frac{2l +1}{4\pi} P_l \left( \hat{{\bm r}}_+ \cdot \hat{{\bm r}}_- \right)
\nonumber
\\
&\times&
\left[
R^{\uparrow}_{n'l}(|{\bm r}_+|) R^{\downarrow}_{nl}(|{\bm r}_-|)
+
R^{\downarrow}_{nl}(|{\bm r}_+|) R^{\uparrow}_{n'l}(|{\bm r}_-|)
\right] d_{lnn'}
,
\label{eq19}
\\
\Phi_{\rm T} \left( {\bm R}, {\bm r}_{\rm rel} \right)
&=&-\frac{1}{\sqrt{2}}\sum_{lnn'} 
\frac{2l +1}{4\pi} P_l \left( \hat{{\bm r}}_+ \cdot \hat{{\bm r}}_- \right)
\nonumber
\\
&\times&
\left[
R^{\uparrow}_{n'l}(|{\bm r}_+|) R^{\downarrow}_{nl}(|{\bm r}_-|)
-
R^{\downarrow}_{nl}(|{\bm r}_+|) R^{\uparrow}_{n'l}(|{\bm r}_-|)
\right] d_{lnn'}
,
\label{eq20}
\end{eqnarray}
where ${\bm r}_\pm={\bm R} \pm {\bm r}_{{\rm rel}}/2$ and $P_l(x)$ is the Legendre polynomial. In particular, when the system possesses the spin rotation symmetry, one find that $d_{lnn'}=d_{ln'n}$ and $R^{\sigma}_{nl}(r)$ is independent of the pseudospin $\sigma$, which immediately gives vanishing $\Phi_{\rm T} \left( {\bm R}, {\bm r}_{\rm rel} \right)$. This indicates the necessity of the broken spin rotation symmetry to induce a non-vanishing $\Phi_{\rm T}\left( {\bm R}, {\bm r}_{\rm {rel}} \right)$.
\par
Besides the Cooper-pair amplitude, the condensate fraction is also a useful quantity to see how many Cooper pairs are Bose-condensed. As usual, the singlet component ($N_{\rm S}^{\rm c}$) and triplet component ($N_{\rm T}^{\rm c}$) of the condensate fraction is given by
\begin{eqnarray}
N_{\rm {S,T}}^{\rm c}=\int d{\bm R} \rho^{\rm c}_{\rm {S,T}} \left( R  \right)
,
\label{eq21}
\end{eqnarray}
where
\begin{eqnarray}
\rho^{\rm c}_{\rm {S,T}} \left( R  \right)=\int d{\bm r}_{\rm rel} |\Phi_{\rm {S,T}} \left( {\bm R}, {\bm r}_{\rm rel} \right)|^2
\label{cfdens}
\end{eqnarray}
is the condensate fraction density of the singlet ($\rho^{\rm c}_{\rm {S}}$) and triplet ($\rho^{\rm c}_{\rm {T}}$) components. For later convenience, we also define the total condensate fraction $N_{\rm {total}}^{\rm c}= N_{\rm {S}}^{\rm c}+N_{\rm {T}}^{\rm c}$, as well as the total condensate fraction density $\rho_{\rm {total}}^{\rm c}= \rho_{\rm {S}}^{\rm c}+\rho_{\rm {T}}^{\rm c}$.
\par
Besides Eqs. (\ref{eq19}) and (\ref{eq20}), it is also useful to classify the pair amplitude in terms of the orbital angular momentum ($L,M$), as
\begin{eqnarray}
\Phi_{LM} \left({\bm R},r_{\rm{rel}} \right) &=& \int d \Omega_{\rm{rel}}  
Y_{LM}\left( \hat{{\bm r}}_{\rm{rel}} \right)
 \left[ 
\Phi_{\rm S} \left({\bm R},{\bm r}_{\rm{rel}} \right) 
+\Phi_{\rm T} \left({\bm R},{\bm r}_{\rm{rel}} \right) 
\right]
,
\label{eq23}
\end{eqnarray}
where the integration is taken over the solid angle with respect to the relative coordinate ${\bm r}_{\rm {rel}}$. The quantities corresponding to Eqs. (\ref{eq21}) and (\ref{cfdens}) are given by, respectively, 
\begin{eqnarray}
N_{L}^c&=& \int d{\bm R} \rho^{\rm c}_{\rm {L}} \left( R  \right)
,
\label{eq24}
\\
\rho^{\rm c}_{\rm {L}} \left( R  \right)&=& \sum_{M=-L}^{L}\int dr_{\rm{rel}}  r_{\rm{rel}}^2 \left| \Phi_{LM} \left({\bm R},r_{\rm{rel}} \right)\right|^2
.
\label{eq25}
\end{eqnarray}
We note that the odd-$L$ (even-$L$) components of $\Phi_{LM}$ are obtained from $\Phi_{\rm T}\left({\bm R} , {\bm r}_{\rm {rel}}\right)$ [$\Phi_{\rm S}\left({\bm R} , {\bm r}_{\rm {rel}}\right)$] in Eq. (\ref{eq23}), because of the well-known antisymmetric property of the wave function of a Fermi pair. Thus, on the viewpoint of the realization a $p$-wave superfluid Fermi gas starting from the $s$-wave superfluid state, the induced pair amplitude $\Phi_{\rm T}$ must involve an $L=1$ ($p$-wave) component $\Phi_{L=1,M}$. In this case, rapidly changing the $s$-wave pairing interaction to an appropriate $p$-wave one, $V_p ({\bm p},{\bm p}')$, by adjusting an external magnetic field from an $s$-wave Feshbach resonance field to a $p$-wave one, we can realize the $p$-wave superfluid state characterized by the $p$-wave superfluid order parameter,
\begin{eqnarray}
\Delta_p \left( {\bm R},{\bm p}\right) = \sum_{\bm p'} V_p \left( {\bm p}, {\bm p}' \right) \Phi_{\rm{T}}\left( {\bm R},  {\bm p}'  \right),
\label{eq22}
\end{eqnarray}
at least just after this manipulation. Here, $\Phi_{\rm {T}}\left( {\bm R},  {\bm p}'  \right)$ is the Fourier-transformed triplet pair amplitude with respect to the relative coordinate ${\bm r}_{\rm {rel}}$. In this paper, thus we mainly consider $\Phi_{L=1,M}\left({\bm R} , {\bm r}_{\rm {rel}}\right)$  in $\Phi_{\rm T}\left({\bm R} , {\bm r}_{\rm {rel}}\right)$.
\par
Before ending this section, we summarize our detailed numerical parameter setting. We take the total number $N=N_\uparrow+N_\downarrow$ of Fermi atoms in a trap as $N=1632$. For the high energy cutoff $\omega_c^{\sigma}$, we set $\omega_c^{\sigma}=501.5\omega_\sigma$, which is much larger than the Fermi energy $\varepsilon_{\rm F}^{\sigma} = ( 6N_\sigma )^{1/3} \omega_\sigma$ in the $\sigma$-component. As usual, we eliminate effects of this cutoff from the theory by introducing the $s$-wave scattering length $a_s$, given by
\begin{eqnarray}
\frac{4\pi a_s}{\bar{m}}=\frac{-U_s}{1-U_s\sum_{\bm p}^{p_{\rm c}} \frac{\bar{m}}{p^2}}
.
\label{eq26}
\end{eqnarray}
Here $p_{\rm c}= \sqrt{2\bar{m} 501.5 \bar{\omega}}$ (where $\bar{\omega} = \sqrt{(\omega_\uparrow^2+\omega_\downarrow^2)/2}$) and  $\bar{m}=2m_{\uparrow}m_{\downarrow}/(m_{\uparrow}+m_{\downarrow})$ is twice the reduced mass. For the temperature, we set $T=0.01\varepsilon_{\rm F}$, where $\varepsilon_{\rm F} = \left( 3N \right)^{1/3} \bar{\omega} \simeq 16.98 \bar{\omega}$. Although we take this small but finite temperature to avoid computational difficulty associated with discrete energy levels in a trap, our results are expected to essentially describe ground state properties of the system. We note that the discreteness of the energy levels still makes our computations difficult in the presence of spin imbalance, even when we take $T=0.01\varepsilon_{\rm F}$. Thus, to avoid this problem, we introduce a small but finite width to one particle energy levels when the system has a finite spin polarization $P=(N_\uparrow-N_\downarrow)/(N_\uparrow+N_\downarrow) \neq 0$. For more details about this manipulation, see the appendix.
\section{$P$-wave Cooper-pair amplitude induced in $s$-wave superfluid Fermi gases}
\begin{figure}
\begin{center}
\includegraphics[%
  width=1.00\linewidth, keepaspectratio]{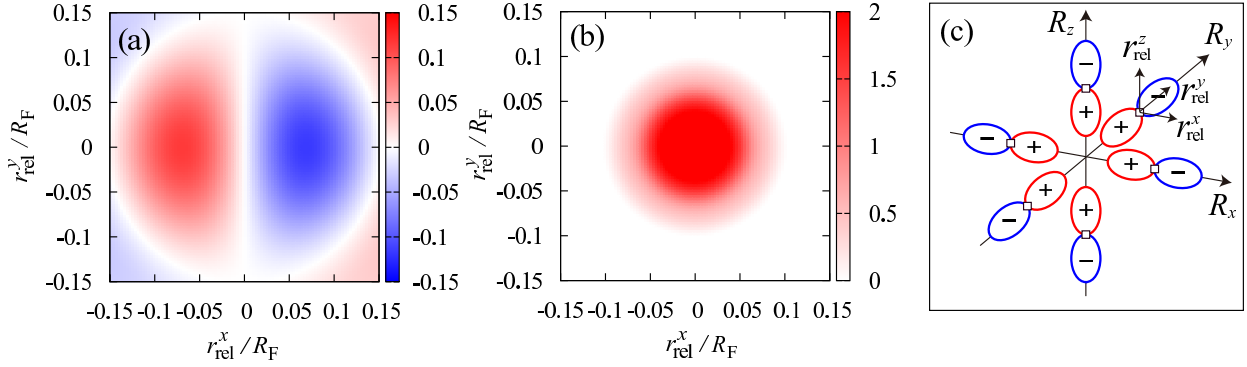}
\end{center}
\caption{(Color online) Calculated (a) spin-triplet component pair amplitude as a function of the relative coordinate $r_{\rm {rel}}^x$ and $r_{\rm {rel}}^y$ in a trapped $s$-wave superfluid Fermi gas with spin-imbalance, $\Phi_{\rm T}\left({\bm R}, {\bm r}_{\rm{rel}} \right)R_{\rm F}^3$, where $R_{\rm F}=\sqrt{2\varepsilon_{\rm F}/(m\bar{\omega})}$ is the Thomas-Fermi radius. (b) Spin-single component $\Phi_{\rm S}\left({\bm R}, {\bm r}_{\rm{rel}} \right)R_{\rm F}^3$. We take $(p_{\rm F} a_s)^{-1}=-0.6$, $P\equiv(N_{\uparrow}-N_{\downarrow})/(N_{\uparrow}+N_{\downarrow}) = 0.2$, $r_{\rm {rel}}^z=0$, and ${\bm R}=(0.8 R_{\rm F}, 0, 0)$. This parameter set is also used in Figs. \ref{fig2} and \ref{fig3}.  (c) Schematic spatial structure of the triplet Cooper-pair amplitude $\Phi_{\rm T} \left({\bm R}, {\bm r}_{\rm{rel}} \right)$. At each center-of-mass position ${\bm R}$ (open square), the ${\bm r}_{\rm{ rel}}$ dependence of $\Phi_{\rm T} \left({\bm R}, {\bm r}_{\rm{rel}} \right)$ is shown.}
\label{fig1}
\end{figure}
Figure 1(a) shows the triplet Cooper-pair amplitude $\Phi_{\rm T} \left({\bm R}, {\bm r}_{\rm{rel}} \right)$ as a function of the relative coordinate ${\bm r}_{\rm {rel}}=(r_{\rm {rel}}^x,r_{\rm {rel}}^y,0)$ induced in a trapped three-dimensional continuum $s$-wave superfluid Fermi gas with spin imbalance. In addition to the singlet component $\Phi_{\rm S} \left({\bm R}, {\bm r}_{\rm{rel}} \right)$ shown in Fig. \ref{fig1}(b), this figure indicates that the Cooper-pair amplitude has the triplet component, when the spatial inversion symmetry and the spin rotation symmetry are simultaneously broken. Although this phenomenon has already been obtained in a lattice model \cite{Endo}, the present result clearly confirms that the presence of background lattice is actually not essential. In this sense, since a spin-imbalanced superfluid Fermi gas has already been realized experimentally \cite{Zwierlein2006,Partridge,Shin}, Fig. \ref{fig1}(a) indicates that the triplet Cooper-pair amplitude has also already been realized in cold Fermi gas system.
\par
In Fig. \ref{fig1}(a), one sees a line node [where $\Phi_{\rm T} \left({\bm R}, {\bm r}_{\rm{rel}} \right)$ vanishes] along the $r_{\rm {rel}}^y$-axis. In this regard, we note that this line node always appears, being perpendicular to ${\bm R}$. [Note that ${\bm R}=(0.8R_{\rm F},0,0)$ in Fig. \ref{fig1}(a), where $R_{\rm F}=\sqrt{2\varepsilon_{\rm F}/(m\bar{\omega})}$ is the Thomas-Fermi radius.] Thus, summarizing $\Phi_{\rm T} \left({\bm R}, {\bm r}_{\rm{rel}} \right)$ of various ${\bm R}$'s, we schematically obtain the overall structure of the triplet Cooper-pair amplitude as shown in Fig. \ref{fig1}(c). In this panel, we briefly note that the triplet pair amplitude vanishes at the trap center ${\bm R}=0$, because the local spatial inversion symmetry holds there.
\par
The reason for the nodal structure seen in Fig. \ref{fig1}(a) can be simply understood by the symmetry consideration. That is, the triplet pair amplitude $\Phi_{\rm T} \left({\bm R}, {\bm r}_{\rm{rel}} \right)$ is generally antisymmetric with respect to the exchange of two Fermi atoms, so that one finds [see also Eq. (\ref{eq17})]
\begin{eqnarray}
\Phi_{\rm T}( {\bm R},-{\bm r}_{{\rm rel}})=-\Phi_{\rm T}({\bm R},{\bm r}_{{\rm rel}}).
\label{eqsym1}
\end{eqnarray}
In addition, since the Hamiltonian in Eq. (\ref{eq2}) is invariant under the spatial rotation around any axis passing through the trap center, $\Phi_{\rm T}( {\bm R},{\bm r}_{{\rm rel}})$ also possesses the same symmetry property. Thus, for the $\pi$-rotation around ${\bm R}$, one finds, for ${\bm r}_{{\rm rel}} \perp {\bm R}$,
\begin{eqnarray}
\Phi_{\rm T}( {\bm R},-{\bm r}_{{\rm rel}})=\Phi_{\rm T}({\bm R},{\bm r}_{{\rm rel}}).
\label{eqsym2}
\end{eqnarray}
Equations (\ref{eqsym1}) and (\ref{eqsym2}) immediately conclude the vanishing $\Phi_{\rm T}( {\bm R},{\bm r}_{{\rm rel}})$ when ${\bm r}_{{\rm rel}} \perp {\bm R}$, giving the nodal line seen in Fig. \ref{fig1}(a).
\par
In our idea explained in the previous section, the induced triplet pair amplitude $\Phi_{\rm T}( {\bm R},{\bm r}_{{\rm rel}})$ shown in Fig. \ref{fig1}(a) is directly used to produce the $p$-wave superfluid order parameter $\Delta_p \left( {\bm R}, {\bm p} \right)$ in Eq. (\ref{eq22}), so that $\Delta_p \left( {\bm R}, {\bm p} \right)$ succeeds to the spatial structure of $\Phi_{\rm T}( {\bm R},{\bm r}_{{\rm rel}})$ shown in Fig \ref{fig1}(c). In this regard, we note that the spatial structure of $\Phi_{\rm T}( {\bm R},{\bm r}_{{\rm rel}})$ is purely determined by the symmetry property of the $s$-wave superfluid state before changing the interaction from the $s$-wave type to the $p$-wave one. Thus, the produced $p$-wave superfluid order parameter $\Delta_p \left( {\bm R}, {\bm p} \right)$ is generally different from that in the ground state of the $p$-wave superfluid phase for a given $p$-wave interaction $V_p ({\bm p},{\bm p}')$. In addition, the $p$-wave interaction is known to cause the particle loss \cite{Gaebler2007,Guririe2007,Gurarie2008,Castin}, so that the realized $p$-wave superfluid phase is inevitably in the non-equilibrium state. However, even when a $p$-wave superfluid Fermi gas is realized by the ordinary approach, where a $p$-wave interaction is finite from the beginning, one cannot avoid the non-equilibrium state. This point would not be a disadvantage of the approach discussed in this paper.
\par
\begin{figure}
\begin{center}
\includegraphics[
  width=1.00\linewidth, keepaspectratio]{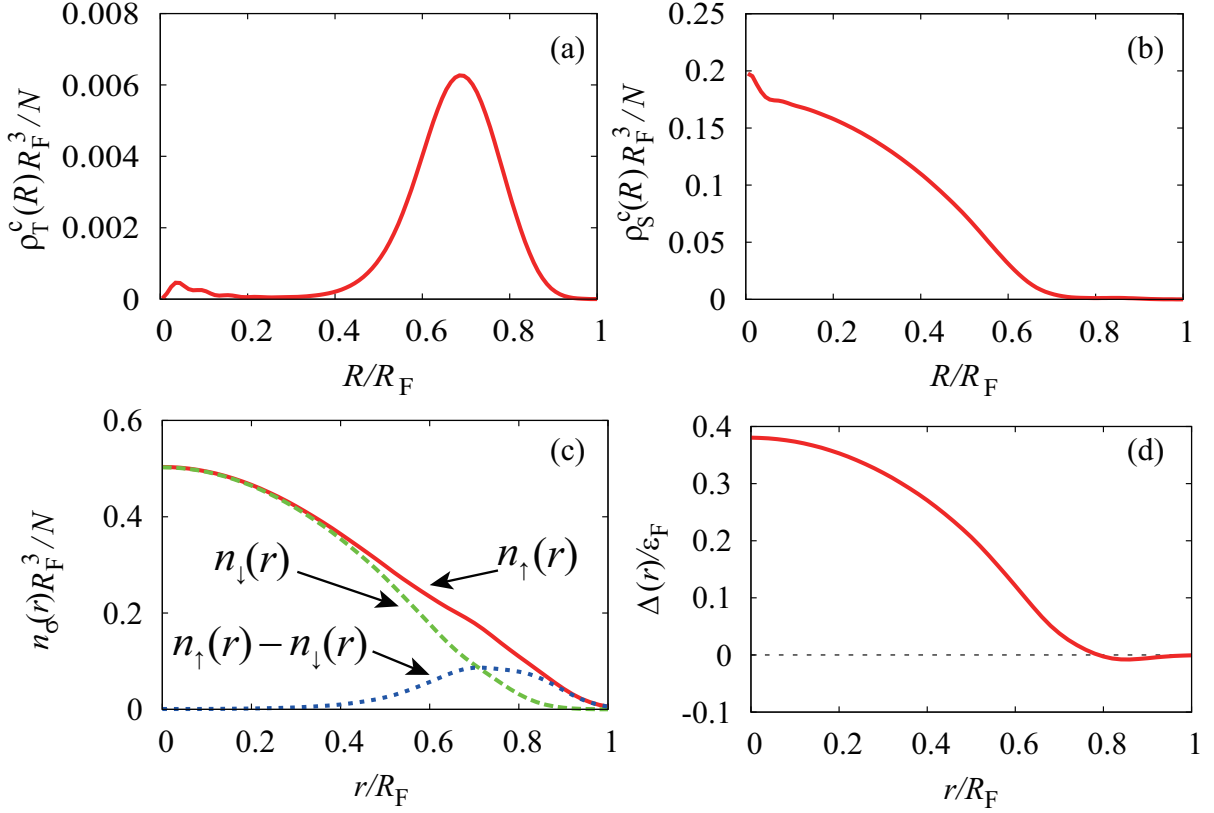}
\end{center}
\caption{(Color online) Calculated (a) spin-triplet component $\rho_{\rm T}^{\rm c}(R)$ and (b) singlet component $\rho_{\rm T}^{\rm c}(R)$ of local condensate fraction in a trapped $s$-wave superfluid Fermi gas with spin imbalance $P \equiv (N_\uparrow-N_\downarrow)/(N_\uparrow+N_\downarrow)=0.2$. (c) $s$-wave superfluid order parameter $\Delta \left( r \right)$. (d) Density profile $n_\sigma (r)$. We take  $\left(p_{\rm F} a_s \right)^{-1}=-0.6$.}
\label{fig2}
\end{figure}
\par
Figure \ref{fig2} (a) shows the triplet component $\rho^{\rm c}_{\rm T} ({\bm R})$ of the local condensate fraction. While the single component $\rho^{\rm c}_{\rm S} ({\bm R})$ has large intensity around the trap center [see Fig. \ref{fig2}(b)], the triplet component $\rho^{\rm c}_{\rm T} ({\bm R})$ is found to take a maximal value near the trap edge ($R \simeq 0.7R_{\rm F}$). For the parity-mixing effect, both the spatial inversion symmetry and the spin rotation symmetry must be broken. While the former condition holds everywhere in a trap except at the trap center (${\bm R}=0$), the latter condition is satisfied around the edge of the gas cloud where the ``local magnetization $M(r)=n_\uparrow(r)-n_\downarrow(r)$" becomes non-vanishing [see Fig. \ref{fig2}(c)]. Because of this, the region where the triplet pair amplitude $\rho^{\rm c}_{\rm T} ({\bm R})$ is  enhanced is almost the same as the spatial region with large local magnetization $M(r)=n_\uparrow(r)-n_\downarrow(r)$, as seen in Figs. \ref{fig2}(a) and (c). We briefly note that this magnetization is directly related to the well-known phase separation phenomenon in a spin-imbalanced superfluid Fermi gas in a trap \cite{Zwierlein2006,Partridge,Shin}.
\par
In Figs. \ref{fig2}(a) and \ref{fig2}(b), one sees that both $\rho^{\rm c}_{\rm T} ({\bm R})$ and $\rho^{\rm c}_{\rm S} ({\bm R})$ are slightly enhanced around the trap center ($R \lesssim 0.05R_{\rm F}$). As the origin for this, we point out the importance of Andreev bound states appearing locally around the trap edge \cite{Ohashitrap}. That is, Fermi atoms feel a trap potential $V(r)=m\omega^2r^2/2$, as well as the off-diagonal pair-potential $\Delta(r)$ shown in Fig. \ref{fig2}(d), so that bound states are formed around the bottom of the combined well of the two. Then, when we simply assume that Andreev bound states are completely localized at $r_0$, their wave functions are written as $\phi({\bm {r}}) \propto \delta (r-r_0)$. In this case, their contribution to the local condensate fraction [$\simeq \int d{\bm r}_{\rm rel} \phi({\bm{R}}-{\bm r}_{\rm rel}/2)\phi({\bm{R}}+{\bm r}_{\rm rel}/2)$] only becomes non-zero at ${\bm R}=0$.
\par
\begin{figure}
\begin{center}
\includegraphics[%
  width=0.50\linewidth, keepaspectratio]{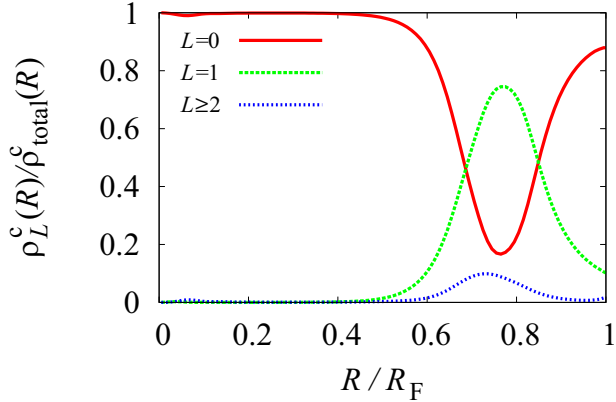}
\end{center}
\caption{(Color online)  Calculated density of condensate fraction $\rho_{L}^{\rm c}(R)$ with the angular momentum $L$, as a function of the center of mass position $R$ of a Cooper pair. We set $(p_{\rm F} a_s)^{-1}=-0.6$, and $P=0.2$. In this case, the total $p$-wave condensate fraction equals $N_{L=1}^{\rm c}/N^{\rm c}_{\rm {total}} \simeq 0.14$.}
\label{fig3}
\end{figure}
For our purpose, the induced triplet pair amplitude shown in Fig. \ref{fig2}(a) must be dominated by the $p$-wave component. To confirm this, we show in Fig. \ref{fig3} the density $\rho_{L}^{\rm c}(R)$ of the condensate fraction in the case of Fig. \ref{fig2}. Indeed, the condensate fraction around $R \simeq 0.8R_{\rm F}$ is dominated by $L=1$ component, that is, the $p$-wave one. This induced $p$-wave component  $\rho_{L=1}^{\rm c}(R)$ is found to amount to about 80\% of the total condensate density around $R \simeq 0.8R_{\rm F}$, which is larger than the magnitude of $s$-wave component density $\rho_{L=0}^{\rm c}(R)$ in this region. Evaluating the condensate fraction of the $p$-wave component, one has $N_{L=1}^{\rm c}=0.14N^{\rm c}_{\rm {total}}$. Thus, this system is found to be useful for the preparation for the $p$-wave pair amplitude {\it without} using a $p$-wave pairing interaction in an ultracold Fermi gas.
\par
\begin{figure}
\begin{center}
\includegraphics[%
  width=0.50\linewidth, keepaspectratio]{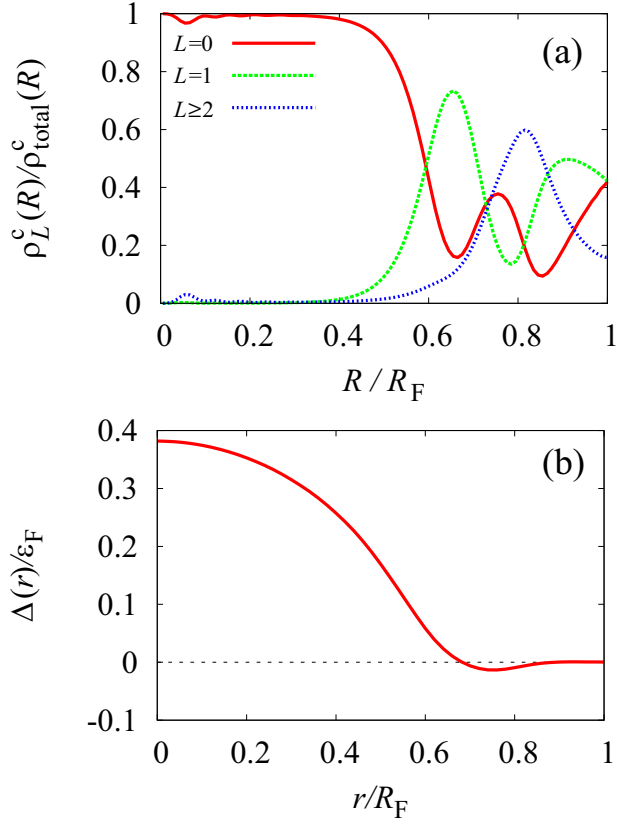}
\end{center}
\caption{(Color online) Same as Fig. \ref{fig3}, when $\left(p_{\rm F} a_s \right)^{-1}=-0.6$ and $P=0.3$.}
\label{fig4}
\end{figure}
In Fig. \ref{fig2}(d), one sees a spatial oscillation of the superfluid order parameter $\Delta(r)$ around the edge of the gas cloud, which is characteristic of the Fulde-Ferrell-Larkin-Ovchinnikov (FFLO) state \cite{FFLO1,FFLO2,FFLO3}. In this regard, although the FFLO state is not necessary for the parity-mixing phenomenon to be realized, it still affects details of the condensate density $\rho_{L}^{\rm c}(R)$, because the spatial oscillation of the FFLO superfluid order parameter also contributes to the local breakdown of the spatial inversion symmetry in addition to the trap potential. In particular, since the wave length of the FFLO order parameter is usually shorter than the characteristic length of the spatial variation by  a trap potential, the FFLO oscillation is expected to induce pair amplitude with high angular momenta, compared to the case without the FFLO state. Indeed, when the FFLO oscillation of the superfluid order parameter $\Delta(r)$ becomes clearer than the case shown in Fig. \ref{fig2} and \ref{fig3}, Fig. \ref{fig4} shows that the condensate density $\rho_{L\ge 2}^{\rm c}(R)$ has a larger value around the edge of gas cloud, as expected.
\par
\begin{figure}
\begin{center}
\includegraphics[%
  width=1.00\linewidth, keepaspectratio]{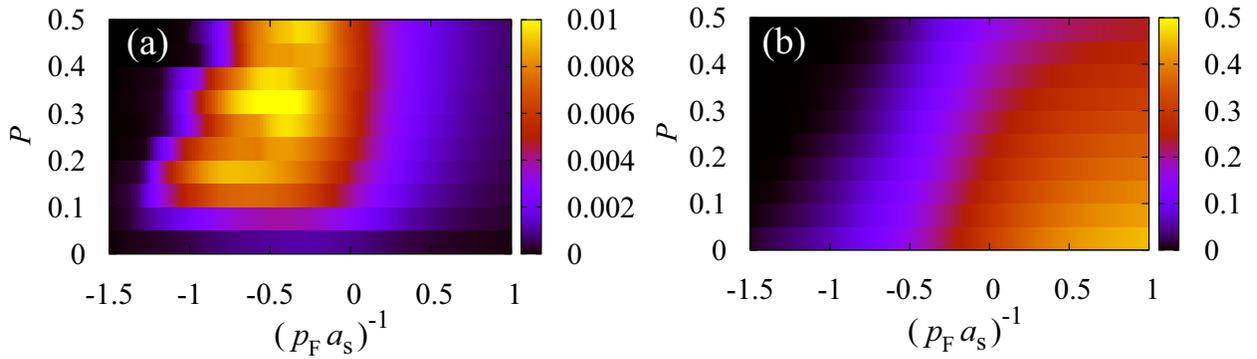}
\end{center}
\caption{(Color online) (a) Calculated condensate fraction $N_{\rm T}^{\rm c}$ of the spin-triplet component. (b) Spin-singlet component $N^c_{\rm S}$. In each panel, the intensity renormalized by the total number of Fermi atoms $N=N_\uparrow+N_\downarrow$.}
\label{fig5}
\end{figure}
To find the region where the parity mixing strongly occurs, we show in Fig. \ref{fig5}(a) the magnitude of the total condensate fraction $N_{\rm T}^{\rm c}$ of the triplet component in the $P$-$(p_{\rm F} a_s)^{-1}$ plane. While the total condensate fraction $N_{\rm S}^{\rm c}$ of the singlet component becomes large with increasing $(p_{\rm F} a_s)^{-1}$ for a given polarization $P$ as shown in Fig. \ref{fig5}(b),  the triplet component $N_{\rm T}^{\rm c}$ is found to take a maximum value in the intermediate-coupling region. This is because, in the strong-coupling BEC regime  atoms form tightly bound spin-singlet $s$-wave molecules, so that the triplet pair amplitude is difficult to appear, leading the suppression of $N_{\rm T}^{\rm c}$. 
\par
\begin{figure}
\begin{center}
\includegraphics[%
  width=1.00\linewidth, keepaspectratio]{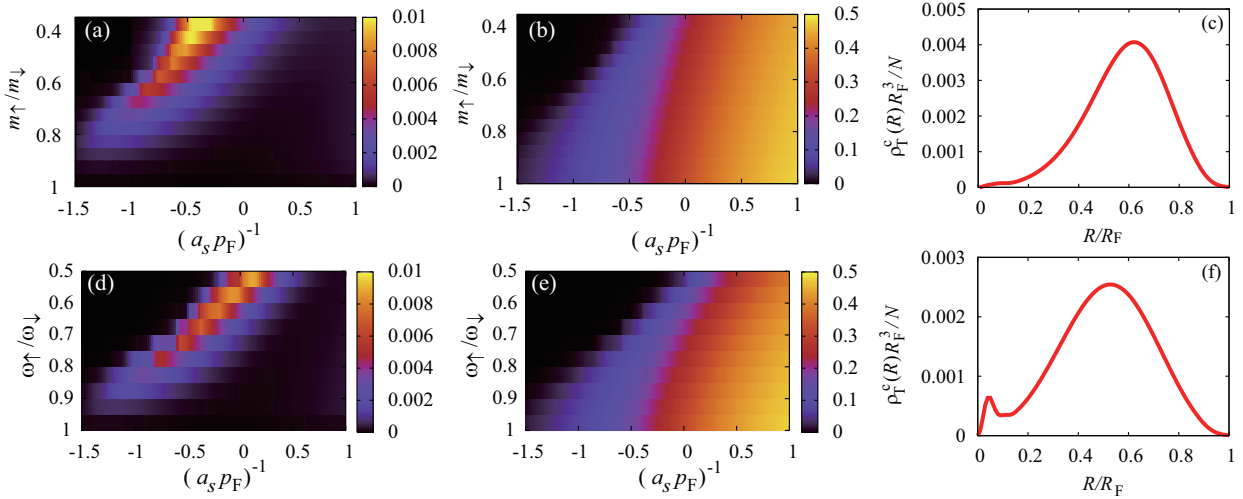}
\end{center}
\caption{(Color online) The upper panels show the case with mass imbalance. In panel (c) we take $m_\uparrow/m_\downarrow=0.5$ and $(p_{\rm F}a_s)^{-1}=-0.5$. The lower ones show the case with spin-dependent trap potential. In panel (f) we take $\omega_\uparrow/\omega_\downarrow=0.8$ and $(a_sp_{\rm F})^{-1}=-0.9$. (a), (d) Spin-triplet component $N_{\rm T}^{c}$ of the total condensate fraction. (b), (e) Spin-singlet component $N_{\rm S}^{c}$ of the total condensate fraction. (c), (f) Triplet component $\rho_{\rm T}^{\rm c}(R)$ of the local condensate fraction.}
\label{fig5_2}
\end{figure}
We point out that the parity-mixing effect also occurs when the spin rotation symmetry is broken by, not spin imbalance, but mass imbalance ($m_\uparrow/m_\downarrow \neq 1$), or a spin-dependent trap potential ($\omega_\uparrow/\omega_\downarrow \neq 1$). As shown in Fig. \ref{fig5_2}, both the cases give similar results to those in the spin-imbalanced case. One crucial difference is that, in these cases, the phase separation between the superfluid region and the normal region does not occur, in contrast to the spin-imbalanced case, so that the triplet Cooper pairs are induced in the wider spatial region, compared to the case of the spin imbalance. [Compare Figs. \ref{fig5_2}(c) and \ref{fig5_2}(f) with \ref{fig2}(a).]
\section{Summary}
To summarize, extending our previous work for the parity-mixing effect in a two-dimensional lattice model \cite{Endo} to a realistic three-dimensional continuum case, we theoretically confirmed that triplet Cooper-pair amplitude $S_z=0$ is induced in a trapped $s$-wave superfluid Fermi gas in the presence of spin imbalance. We showed that the induced triplet pair amplitude has a unique spatial structure coming from the spatial rotation symmetry of the system. We also  numerically evaluated how large the triplet pair amplitude involves the $p$-wave component. In addition, we also pointed out that a similar effect also occurs when the spin rotation symmetry is broken by mass imbalance or spin-dependent trap potential. 
\par
The existence of $p$-wave pairs does not immediately means the realization of the $p$-wave superfluid state, because the symmetry of the superfluid state is still characterized by the $s$-wave {\it superfluid order parameter}. However, in an ultracold Fermi gas, one can rapidly switch the interaction from an $s$-wave one to a $p$-wave one by using Feshbach resonance. Thus, combining this technique with the parity-mixing effect discussed in this paper, one may reach the $p$-wave superfluid state, at least just after turning on the $p$-wave interaction. In this regard, we emphasize that a trapped $s$-wave superfluid with spin imbalance has already been realized in a $^6$Li Fermi gas \cite{Zwierlein2006,Partridge,Shin}, so that $p$-wave Cooper-pair amplitude is expected to have also already been induced then. Since experiments toward the realization of a $p$-wave superfluid are facing various difficulties, such as three-body loss \cite{Guririe2007,Gurarie2008,Castin}, as well as dipolar relaxation \cite{Gaebler2007}, our results would provide an alternative approach for this exciting challenge.
\par
\begin{acknowledgements}
\par
We thank Y. Endo for useful discussions. This work was supported by KiPAS project in Keio University. DI was supported by JSPS KAKENHI (No.JP16K17773). RH was supported by a Grant-in-Aid for JSPS fellows. YO was supported by Grant-in-Aid for Scientific research from MEXT and JSPS in Japan (No.JP15H00840, No.JP15K00178, No.JP16K05503). 
\end{acknowledgements}
\appendix
\section{Detailed computations in the presence of spin imbalance}
\label{apA}
\begin{figure}
\begin{center}
\includegraphics[%
  width=1.00\linewidth, keepaspectratio]{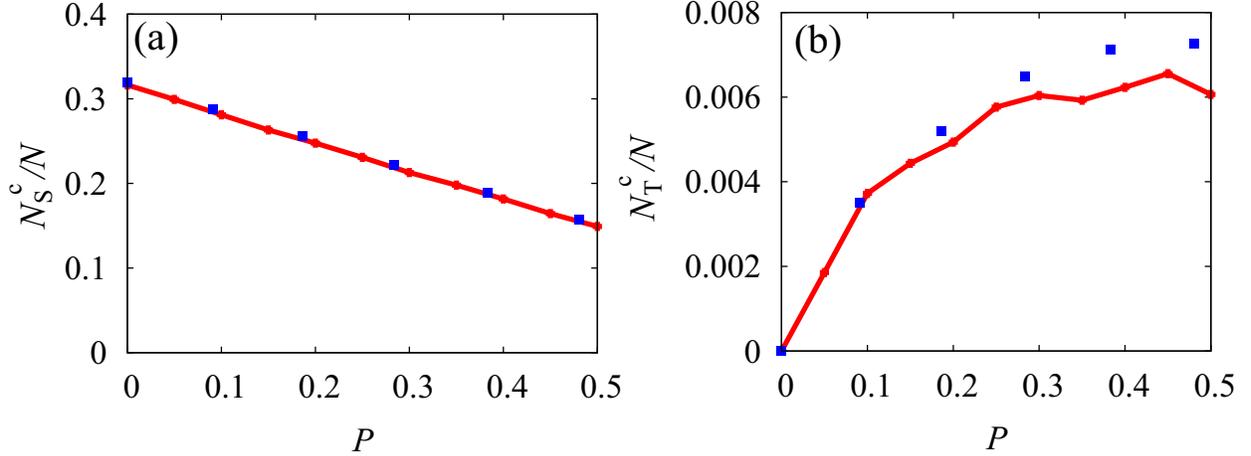}
\end{center}
\caption{(Color online) Calculated condensate fraction in a trapped $s$-wave superfluid Fermi gas with spin imbalance. (a) spin-singlet component $N^{\rm c}_{\rm S}$. (b) spin-triplet component $N^{\rm c}_{\rm T}$. We take $(p_{\rm F} a_s)^{-1}=0$. The solid squares are the results with $\gamma=0$ and the solid line shows the results when $\gamma=0.05\varepsilon_{\rm F}$. In obtaining the results with $\gamma=0$, we tuned the spin polarization $P=(N_\uparrow-N_\downarrow)/(N_\uparrow+N_\downarrow)$ by varying $N_\downarrow$ for $N_\uparrow=816$.}
\label{fig6}
\end{figure}
In the presence of spin imbalance ($P \neq 0$), setting $T=0.01\varepsilon_{\rm F}$ is not enough to completely eliminate the computational difficulty coming from the discrete energy levels in a trap. Thus in this paper, we introduce a small but finite width $\gamma=0.05\varepsilon_{\rm F}$ to each eigenenergy level at $E_j^{l}$. This is achieved by replacing $d_{lnn'}$, $\eta^{\uparrow}_{lnn'}$ and $\eta^{\downarrow}_{lnn'}$ in Eqs. (\ref{eq9}) and (\ref{eq10}) by
\begin{eqnarray}
d_{lnn'}&=&\sum_{i=1}^{2(N_l+1)} W^{l}_{n+N_l+1,i}W^{l}_{n',i} 
\frac{1}{\pi} \int d\omega
\frac{\gamma}{\left( \omega - E^l_i \right)^2 + \gamma^2}
f \left(\omega \right)
,
\label{eqa1}
\\ 
\eta^\uparrow_{lnn'}&=&\sum_{i=1}^{2(N_l+1)} W^{l}_{n,i}W^{l}_{n',i}
\frac{1}{\pi} \int d\omega
\frac{\gamma}{\left( \omega - E^l_i \right)^2 + \gamma^2}
f \left(\omega \right)
,
\label{eqa2}
\\ 
\eta^\downarrow_{lnn'}&=&\sum_{i=1}^{2(N_l+1)} W^{l}_{n+N_l+1,i}W^{l}_{n'+N_l+1,i} 
\frac{1}{\pi} \int d\omega
\frac{\gamma}{\left( \omega - E^l_i \right)^2 + \gamma^2}
\left[ 1- f \left( \omega \right) \right]
.
\label{eqa3}
\end{eqnarray}
As shown in Fig. \ref{fig6}, although this manipulation slightly lowers the magnitude of the triplet condensate fraction $N_{\rm T}^{\rm c}$ when the spin polarization becomes large to some extent, using Eqs. (\ref{eqa1})-(\ref{eqa3}) does not affect the essence of the parity-mixing effect, giving a non-vanishing value of the triplet condensate fraction in an $s$-wave superfluid Fermi gas.
\par

\par


\begin{thebibliography}{99}
\bibitem{Endo} Y. Endo, D. Inotani, R. Hanai, and Y. Ohashi Phys. Rev. A \textbf{92}, 023610 (2015).
\bibitem{Regal2003} C. A. Regal, C. Ticknor, J. L. Bohn, and D. S. Jin, Phys. Rev. Lett. {\bf 90}, 053201 (2003).
\bibitem{Ticknor2004} C. Ticknor, C. A. Regal, D. S. Jin, and J. L. Bohn, Phys. Rev. A {\bf 69}, 042712 (2004).
\bibitem{Salomon2004} J. Zhang, E. G. M. van Kempen, T. Bourdel, L. Khaykovich, J. Cubizolles, F. Chevy, M. Teichmann, L. Tarruell, S. J. J. M. F. Kokkelmans, and C. Salomon, Phys. Rev. A {\bf 70}, 030702 (2004).
\bibitem{Gunter2005} K. G\"unter, T. St\"oferle, H. Moritz, M. K\"ohl, and T. Esslinger, Phys. Rev. Lett. {\bf 95}, 230401 (2005).
\bibitem{Gaebler2007} J. P. Gaebler, J. T. Stewart, J. L. Bohn, and D. S. Jin, Phys. Rev. Lett. {\bf 98}, 200403 (2007).
\bibitem{Ketterle2005} C. H. Schunck, M. W. Zwierlein, C. A. Stan, S. M. F. Raupach, W. Ketterle, A. Simoni, E. Tiesinga, C. J. Williams, and P. S. Julienne, Phys. Rev. A {\bf 71}, 045601 (2005).
\bibitem{Mukaiyama2013} T. Nakasuji, J. Yoshida, and T. Mukaiyama, Phys. Rev. A {\bf 88}, 012710 (2013).
\bibitem{Guririe2007} J. Levinsen, N. R. Cooper, and V. Gurarie, Phys. Rev. Lett. {\bf 99}, 210402 (2007).
\bibitem{Gurarie2008} J. Levinsen, N. R. Cooper, and V. Gurarie, Phys. Rev. A {\bf 78}, 063616 (2008).
\bibitem{Castin} M. Jona-Lasinio, L. Pricoupenko, and Y. Castin, Phys. Rev. A {\bf 77}, 043611 (2008).
\bibitem{Ho2005} T.-L. Ho and R. B. Diener, Phys. Rev. Lett. {\bf 94}, 090402 (2005).
\bibitem{Ohashi2005} Y. Ohashi, Phys. Rev. Lett. {\bf 94}, 050403 (2005).
\bibitem{Inotani2012} D. Inotani, R. Watanabe, M. Sigrist, and Y. Ohashi, Phys. Rev. A {\bf 85}, 053628 (2012).
\bibitem{Inotani2015} D. Inotani and Y. Ohashi, Phys. Rev. A {\bf 92}, 063638 (2015).
\bibitem{Regal} C. A. Regal, M. Greiner, and D. S. Jin, Phys. Rev. Lett. \textbf{92}, 040403 (2004).
\bibitem{Zwierlein} M. W. Zwierlein, C. A. Stan, C. H. Schunck, S. M. F. Raupach, A. J. Kerman, and W. Ketterle, Phys. Rev. Lett. {\bf 92}, 120403 (2004).
\bibitem{Kinast} J. Kinast, S. L. Hemmer, M. E. Gehm, A. Turlapov, and J. E. Thomas, Phys. Rev. Lett. \textbf{92}, 150402 (2004).
\bibitem{Bartenstein} M. Bartenstein, A. Altmeyer, S. Riedl, S. Jochim, C. Chin, J. H. Denschlag, and R. Grimm, Phys. Rev. Lett. {\bf 92}, 203201 (2004).
\bibitem{Horikoshi} M. Horikoshi, S. Nakajima, M. Ueda, and T. Mukaiyama, Science \textbf{327}, 442 (2010).
\bibitem{Tsuchiya} S. Tsuchiya, R. Watanabe, and Y. Ohashi, Phys. Rev. A {\bf 82}, 033629 (2010), {\it ibid}, \textbf{84}, 043647 (2011).
\bibitem{Watanabe} R. Watanabe, S. Tsuchiya, and Y. Ohashi, Phys. Rev. A \textbf{86}, 063603 (2012).
\bibitem{Haussmann} R. Haussmann and W. Zwerger, Phys. Rev. A \textbf{78}, 063602 (2008).
\bibitem{Bulgac} A. Bulgac, J. E. Drut, and P. Magierski, Phys. Rev. Lett. \textbf{99}, 120401 (2007). 
\bibitem{deGennes} P. G. de Gennes, {\it Superconductivity of Metals and Alloys} (Addison-Wesley, New York, 1989).
\bibitem{Ohashitrap} Y. Ohashi and A. Griffin, Phys. Rev. A {\bf 72}, 063606 (2005), {\it ibid}, \textbf{72}, 013601 (2005).
\bibitem{Zwierlein2006} M. W. Zwierlein, A. Schirotzek, C. H. Schunck, and W. Ketterle, Science, {\bf 311}, 492 (2006).
\bibitem{Partridge} G. B. Partridge, W. Li, R. I. Kamar, Y.-A. Liao, and R. G. Hulet, Science {\bf 311}, 503 (2006).
\bibitem{Shin} Y. Shin, C. H. Schunck, A. Schirotzek, and W. Ketterle, Nature (London) \textbf{451}, 689 (2008).
\bibitem{Kashimura} T. Kashimura, R. Watanabe, and Y. Ohashi, Phys. Rev. A \textbf{86}, 043622 (2012), {\it ibid}, \textbf{89}, 013618 (2014). 
\bibitem{Taglieber2008} M. Taglieber, A.-C. Voigt, T. Aoki, T. W. H{\"a}nsch, and K. Dieckmann, Phys. Rev. Lett. {\bf 100}, 010401 (2008).
\bibitem{Wille2008} E. Wille, F. M. Spiegelhalder, G. Kerner, D. Naik, A. Trenkwalder, G. Hendl, F. Schreck, R. Grimm, T. G. Tiecke, J. T. M. Walraven, S. J. J. M. F. Kokkelmans, E. Tiesinga, and P. S. Julienne, Phys. Rev. Lett. {\bf 100}, 053201 (2008).
\bibitem{Spiegelhalder2009} F. M. Spiegelhalder, A. Trenkwalder, D. Naik, G. Hendl, F. Schreck, and R. Grimm, Phys. Rev. Lett. {\bf 103}, 223203 (2009).
\bibitem{Lin2006} G.-D. Lin, W. Yi and L. -M. Duan, Phys. Rev. A {\bf 74}, 031604(R) (2006).
\bibitem{Hanai2013} R. Hanai, T. Kashimura, R. Watanabe, D. Inotani, and Y. Ohashi, Phys. Rev. A {\bf 88}, 053621 (2013).
\bibitem{Hanai2014} R. Hanai and Y. Ohashi, Phys. Rev. A {\bf 90}, 043622 (2014).
\bibitem{revSOC} For a review, see, J. Dalibard, F. Gerbier, G. Juzeli${\bar {\rm u}}$nas, and P. \"Ohberg, Rev. Mod. Phys. \textbf{83}, 1523 (2011). 
\bibitem{Rb-SOC1} Y.-J. Lin, R. L. Compton, A. R. Perry, W. D. Phillips, J. V. Porto, and I. B. Spielman, Phys. Rev. Lett. \textbf{102}, 130401 (2009). 
\bibitem{Rb-SOC2} Y.-J. Lin, R. L. Compton, K. Jim\'{e}nez-Garc\'{i}a, J. V. Porto, and I. B. Spielman, Nature \textbf{462}, 628 (2009). 
\bibitem{Rb-SOC3} Y.-J. Lin, R. L. Compton, K. Jim\'{e}nez-Garc\'{i}a, W. D. Phillips, J. V. Porto, and I. B. Spielman, Nat. Phys. \textbf{7}, 531 (2011). 
\bibitem{Rb-SOC4} Y.-J. Lin, K. Jim\'{e}nez-Garc\'{i}a, and I. B. Spielman, Nature \textbf{471}, 83 (2011). 
\bibitem{Li-SOC} L. W. Cheuk, A. T. Sommer, Z. Hadzibabic, T. Yefsah, W. S. Bakr, and M. W. Zwierlein, Phys. Rev. Lett. \textbf{109}, 095302 (2012).
\bibitem{K-SOC} P. Wang, Z.-Q. Yu, Z. Fu, J. Miao, L. Huang, S. Chai, H. Zhai, and J. Zhang, Phys. Rev. Lett. \textbf{109}, 095301 (2012).
\bibitem{coldSOC5} J. D. Sau, R. Sensarma, S. Powell, I. B. Spielman, S. DasSarma, Phys. Rev. B \textbf{83}, 140510(R) (2011).  
\bibitem{coldSOC6} B. M. Anderson, I. B. Spielman, and G. Juzeli\={u}nas, Phys. Rev. Lett. \textbf{111}, 125301 (2013). 
\bibitem{Yamaguchi} T. Yamaguchi and Y.Ohashi, Phys. Rev. A \textbf{92}, 013615 (2015).
\bibitem{Hu2011} L. Jiang, X. J. Liu, H. Hu, and H. Pu, Phys. Rev. A {\bf 84}, 063618 (2011).
\bibitem{2B1} H. Hu, L. Jiang, X.-J. Liu, and H. Pu, Phys. Rev. Lett. {\bf 107}, 195304 (2011). 
\bibitem{Vollhardt}  D. Vollhardt and P. W\"olfle, {\it The Superfluid phase of Helium 3} (Taylor and Francis, London, 1990).
\bibitem{FFLO1} P. Fulde, and R. A. Ferrell, Phys. Rev. {\bf 135}, A550 (1964).
\bibitem{FFLO2} A. I. Larkin, and Yu. N. Ovchinnikov, Sov. Phys. JETP {\bf 20}, 762 (1965).
\bibitem{FFLO3} S. Takada, and T. Izuyama, Prog. Theor. Phys. {\bf 41}, 635 (1969).
\end{thebibliography}
\end{document}